\documentclass[twoside]{LCWS11}
\usepackage[latin1]{inputenc}
\usepackage[dvips]{graphicx,epsfig,color}
\usepackage{wrapfig,rotating}
\usepackage{amssymb,amsmath,array}
\usepackage{cite}
\def\ga{\mathrel{\raise.3ex\hbox{$>$\kern-.75em\lower1ex\hbox{$\sim$}}}}
\def\la{\mathrel{\raise.3ex\hbox{$<$\kern-.75em\lower1ex\hbox{$\sim$}}}}

\begin{document}

\title{
$h\gamma\gamma$ coupling in Higgs Triplet Model} 
\author{A.~Arhrib$^*$${}^{1,2}$, R. Benbrik${}^{2,3, 4}$,  M.~Chabab${}^2$,
        G.~Moultaka${}^{5,6}$ and L.~Rahili${}^2$
{\thanks{This work was supported by Programme Hubert Curien,
Volubilis, AI n0: MA/08/186 as well as the
LIA (International Laboratory for Collider Physics-ILCP).}}
\vspace{.3cm}\\
1- Facult\'e des Sciences et Techniques, Tanger, Morocco.\\
2- LPHEA, FSSM, Universit\'e Cadi-Ayyad, Marrakech, Morocco. \\
3- Facult\'e Polydisciplinaire, Universit\'e
Cadi Ayyad, Sidi Bouzid, Safi, Morocco.\\
4-Instituto de Fisica de Cantabria (CSIC-UC),
Santander, Spain\\
5-Universit\'e Montpellier 2, Laboratoire Charles
Coulomb UMR5221, F-34095 Montpellier, France.\\
6-CNRS, Laboratoire Charles Coulomb UMR 5221,  F-34095
  Montpellier, France.
}

\maketitle

\begin{abstract}
We investigate Higgs boson decay
into two photons in the type-II seesaw model. The rate of
$h\to \gamma\gamma$ gets suppressed/enhanced
in this model compared to the Standard Model
(SM) due to the presence of the singly and doubly charged Higgs $H^\pm$ and
$H^{\pm\pm}$.
%
\end{abstract}

\section{Introduction}
One of the main goals of the LHC is the search for the scalar Higgs bosons
and the exploration of the mechanism which is responsible for the
electroweak symmetry breaking.
However, in order to establish the Higgs mechanism as the correct one for
electroweak symmetry breaking, we need to measure the Higgs couplings
to fermions and to gauge bosons as well as the self-interaction of
Higgs bosons. If those measurements are precise enough,
they can be helpful in discriminating between the models
through their sensitivity to quantum correction effects. \\
It is now well known, that at the LHC the branching ratio
$h\to \gamma\gamma$ can be extracted
with a precision of the order 15\% at the high luminosity option of LHC
 \cite{Weiglein:2004hn} while at the International Linear Collider (ILC)
it can be exctacted with about 23-25\% for Higgs mass arround 120 GeV
 \cite{Djouadi:2007ik}. In the case where the $\gamma\gamma$ option of ILC
is available, with the 1\% accuracy measurement of
the branching ratio of $h\to b\bar{b}$ at ILC,
the width $\Gamma(h\to \gamma\gamma)$ can be determined with 2\%
accuracy from $\gamma\gamma\to h^*\to b\bar{b}$ process,
for Higgs mass of 120 GeV \cite{Djouadi:2007ik}. Therefore, precise
calculation of $\Gamma(h\to\gamma\gamma)$ within different beyond Standard
Model (SM) is highly needed.

Recently, the ATLAS and CMS experiments have already probed
the Higgs boson in the mass range  $110$--$600$ GeV, and excluded a Standard
Model (SM) Higgs  in  the range $141$--$476$ GeV at the $95 \%$C.L.
through a combined analysis of all decay channels and up to
$\sim 2.3 {\rm fb}^{-1}$ integrated luminosity per experiment,
\cite{ATLAS-CONF-2011-163}.
Very recently, CMS and ATLAS exclude with $4.9 {\rm fb}^{-1}$ datasets
 $1$ to $2$--$3$ times the SM diphoton cross-section
at the  $95 \%$C.L. in most of the mass range $110$--$130$~GeV,
and report an excess of events around $123$--$127$~GeV in
the diphoton channel, corresponding to an exclusion of
$3$ and $4$ times the SM cross-section respectively for CMS
\cite{CMS-PAS-HIG-11-030} and ATLAS \cite{ATLAS-CONF-2011-161}.
Furthermore, they exclude a SM Higgs in small,
though different, portions of this mass range,
$(112.7)114$--$115(.5)$~GeV for ATLAS and $127$--$131$ GeV for CMS,
at the $95 \%$C.L.


The SM Higgs sector, extended by one  weak gauge
triplet of scalar fields (hereafter dubbed DTHM),
is a very promising setting to account
for neutrino masses through the so-called type II seesaw mechanism.
This Higgs sector, containing two CP-even, one CP-odd,
one charged and one doubly-charged Higgs scalars,
 can be tested directly at the LHC or ILC, provided that
the Higgs triplet mass scale $M_{\Delta}$ and the soft
lepton-number violating mass parameter $\mu$ are of order or below the
weak-scale \cite{Perez:2008ha,Arhrib:2011uy}.
Moreover, in most of the parameter space  [and apart from an extremely narrow
region of $\mu$], one of the two  CP-even Higgs scalars is
generically essentially SM-like and the other an almost
decoupled triplet, irrespective of their relative masses, \cite{Arhrib:2011uy}.

It follows that
if all the Higgs sector of the model is accessible to the LHC or ILC,
one expects a neutral Higgs state with
cross-sections very close to the SM in all Higgs production
and decay channels to leading electroweak order,
except for the di-photon channel. Indeed, in the latter channel,
loop effects of the other
Higgs states can lead to substantial enhancements which can
then be readily analyzed in the light of the experimental
exclusion limits as argued above.

In this paper we will analyze the decay
$h\to \gamma\gamma$ in the framework of DTHM.
This effect will mainly come from singly and doubly charged Higgs boson
contributions.  We will show  that  DTHM
 can account for the excess in the di-photon cross-section
 reported  by ATLAS/CMS, but it can also account for a deficit in
 the di-photon cross-section without modifying the gluon
   fusion rate as well as the other channels like
$h\to b\bar{b}, \tau^+\tau^-, WW^*,ZZ^*$.
\section{Higgs sector of DTHM}
The scalar sector of the DTHM model consists of the
standard Higgs doublet $H$ and a colorless Higgs triplet $\Delta$ with
hypercharge $Y_H=1$ and $Y_\Delta=2$ respectively.
Their matrix representation are given by:
\begin{eqnarray}
\Delta &=\left(
\begin{array}{cc}
\delta^+/\sqrt{2} & \delta^{++} \\
\delta^0 & -\delta^+/\sqrt{2}\\
\end{array}
\right) \qquad {\rm and} \qquad H=\left(
                    \begin{array}{c}
                      \phi^+ \\
                      \phi^0 \\
                    \end{array}
                  \right)
\end{eqnarray}
The most general $SU(2)_{L}\times U(1)_{Y}$ gauge invariant
renormalizable potential $V(H, \Delta)$ is given by
\cite{Arhrib:2011uy, Perez:2008ha}:
\begin{eqnarray}
V&=&-m_H^2{H^\dagger{H}}+\frac{\lambda}{4}(H^\dagger{H})^2+
M_\Delta^2Tr(\Delta^{\dagger}{\Delta}) +
\lambda_1(H^\dagger{H})Tr(\Delta^{\dagger}{\Delta})\label{eq:VDTHM} \\ & + &
\lambda_2(Tr\Delta^{\dagger}{\Delta})^2
+\lambda_3Tr(\Delta^{\dagger}{\Delta})^2
+ \lambda_4{H^\dagger\Delta\Delta^{\dagger}H}+
[\mu(H^T{i}\tau_2\Delta^{\dagger}H)+hc]\nonumber
\end{eqnarray}
Once EWSB takes place, the neutral components of the
Higgs doublet and Higgs triplet acquire vacuum
expectation values \cite{Arhrib:2011uy}.
The DTHM is fully specified by seven independent parameters which we will
take: $\lambda$, $\lambda_{i=1...4}$, $\mu$  and $v_t$.
These parameters respect a set of dynamical
constraints originating from the potential, particularly
perturbative unitarity and boundedness from below constraints
\cite{Arhrib:2011uy}.
The model spectrum contains seven physical Higgs states:
a pair of CP even states $(h, H)$ with $m_h<m_H$, 
one CP odd Higgs boson $A$, one simply charged Higgs $H^\pm$ 
and one doubly charged state $H^{\pm\pm}$.

The mass of the SM-Higgs like $h$ is fixed more or less by $\lambda$ parameter
while the charged Higgs state masses, given below,
will depend strongly on $\lambda_4$ and $\mu$:
\begin{eqnarray}
m_{H^\pm}^2&=&\frac{(v_d^2+2 v_t^2)[2\sqrt{2}\mu- \lambda_4 v_t]}{4v_t}
\label{eq:mHpm} \\
m_{H^{\pm\pm}}^2&=&\frac{\sqrt{2}\mu{v_d^2}- \lambda_4v_d^2v_t-2\lambda_3v_t^3}{2v_t}  \label{eq:mHpmpm}
\end{eqnarray}
For a recent and comprehensive study of the DTHM,
in particular concerning the distinctive properties of the mixing
angle between the neutral components of the doublet and triplet  Higgs fields
and its correlation with $\mu$ parameter,
we refer to \cite{Arhrib:2011uy}.

\section{$h\to \gamma\gamma$}
The low SM Higgs mass region, $[110,140]$ GeV, is the most
challenging for LHC searches. In this mass regime, the main search
channel through the rare decay into a pair of photons as well as
the decay into $\tau^+ \tau^-$.
The Higgs decay into two photons is only possible through loops.
The theoretical predictions for its decay rate is well known in
the SM for long time. Following the ref~\cite{Arhrib:2011uy},
 we will see how singly charged ($H^\pm$) and doubly charged ($H^{\pm\pm}$)
Higgs states of the DTHM could enhance or suppress the two photons decay rate.
Furthermore, since one or the other of the two CP-even neutral Higgs
bosons $h, H$ present in the DTHM
can behave as a purely SM-like Higgs depending on the $\mu$ parameter
which fixes the regime under consideration (see \cite{Arhrib:2011uy}),
we will refer to the SM-like state generically as $H$ in the following.
In the present paper, we will concentrate only on $h$ as the SM-like Higgs,
for the other case we refer to \cite{Arhrib:2011vc} for details.

The decay $h\to \gamma\gamma$ is mediated at 1-loop level by the
virtual exchange of the SM fermions, the SM gauge bosons and the new 
charged Higgs states. Detailed analytic expression for the partial width
$\Gamma(h\to \gamma\gamma)$ can be found in \cite{Arhrib:2011vc}.
Note that the structure of the $H^\pm$ and $H^{\pm \pm}$
contributions is the same except for the fact that the  $H^{\pm \pm}$
contribution is enhanced by a relative factor four in the amplitude
since $H^{\pm  \pm}$ has an electric charge of $\pm 2$ units. For the
following discussion we denote $A_0^{h}(H^\pm)$, $A_0^h(H^{\pm\pm})$ and 
$A_1^h(W)$ respectively the singly charged Higgs, doubly charged Higgs and the W
boson contributions to $h\to \gamma\gamma$.  \\
The coupling of the SM-like higgs to the new charged Higgs states is given by:
\begin{eqnarray}
g_{h H^{++}H^{--}}  \approx   - \bar{\epsilon}
\lambda_1v_d \label{eq:gcalHHpp},\quad
g_{h H^+H^-} \approx
- \bar{\epsilon} (\lambda_{1} + \frac{\lambda_{4}}{2}) v_d \label{eq:gcalHHp}
\end{eqnarray}
where $\bar{\epsilon}$ is the sign of
$s_\alpha$, the mixing between doublet and triplet components,
in the convention where $c_\alpha$ is always positive.

As well known, the decay width of
$h \to \gamma\gamma$ in the SM is dominated by the W loops which can
also interfere destructively with the subdominant top contribution.
In the DTHM, the signs of the couplings $g_{hH^+H^-}$
and $g_{h H^{++}H^{--}}$, and thus those of the  $H^\pm$ and $H^{\pm \pm}$ contributions to $\Gamma(h \rightarrow\gamma\gamma)$,
are fixed respectively by the signs of $2\lambda_1+\lambda_4$ and $\lambda_1$,
Eqs.(\ref{eq:gcalHHpp}, \ref{eq:gcalHHp}).
However, the combined perturbative unitarity and
potential boundedness from below (BFB) constraints
derived in \cite{Arhrib:2011uy} confine $\lambda_1, \lambda_4$ to
small regions. For instance, in the case of vanishing
$\lambda_{2,3}$, $\lambda_1$ is forced to be positive while
$\lambda_4$ can have either signs but still with bounded values
of $|\lambda_4|$ and $|2\lambda_1+\lambda_4|$. Moreover, since
we are considering scenarios where $\mu \sim {\cal O}(v_t)$,
negative values of $\lambda_4$ can be
favored by the experimental bounds on the
(doubly)charged Higgs masses, Eqs.~(\ref{eq:mHpm},\ref{eq:mHpmpm}).
For definiteness we stick in the following to $\lambda_1 >0$,
although the sign of $\lambda_1$ can be relaxed if
$\lambda_{2,3}$ are non-vanishing. Also in the considered mass range
for $h, H^\pm$ and $H^{\pm \pm}$, the charged state contributions
$A_{0}^{h}(H^\pm,H^{\pm\pm})$ are real-valued and take positive values
in the range $ 0.3 - 1$.  An increasing value of $\lambda_1$
will thus lead to contributions of $H^\pm$ and $H^{\pm \pm}$
that are constructive among each other but destructive with
respect to the sum of $W$ boson and top quark contributions.
[Recall that for the W contributions, ${\cal R}e A_{1}^{h}(W)$
takes negative values in the range $-12$ to $-7$.]
As we will see in the next section, this can either reduce tremendously the
branching ratio into di-photons, or increase it by an amount
that can be already constrained by the present ATLAS/CMS results.



\begin{wrapfigure}{r}{0.5\columnwidth}
\centerline{\includegraphics[width=0.45\columnwidth]{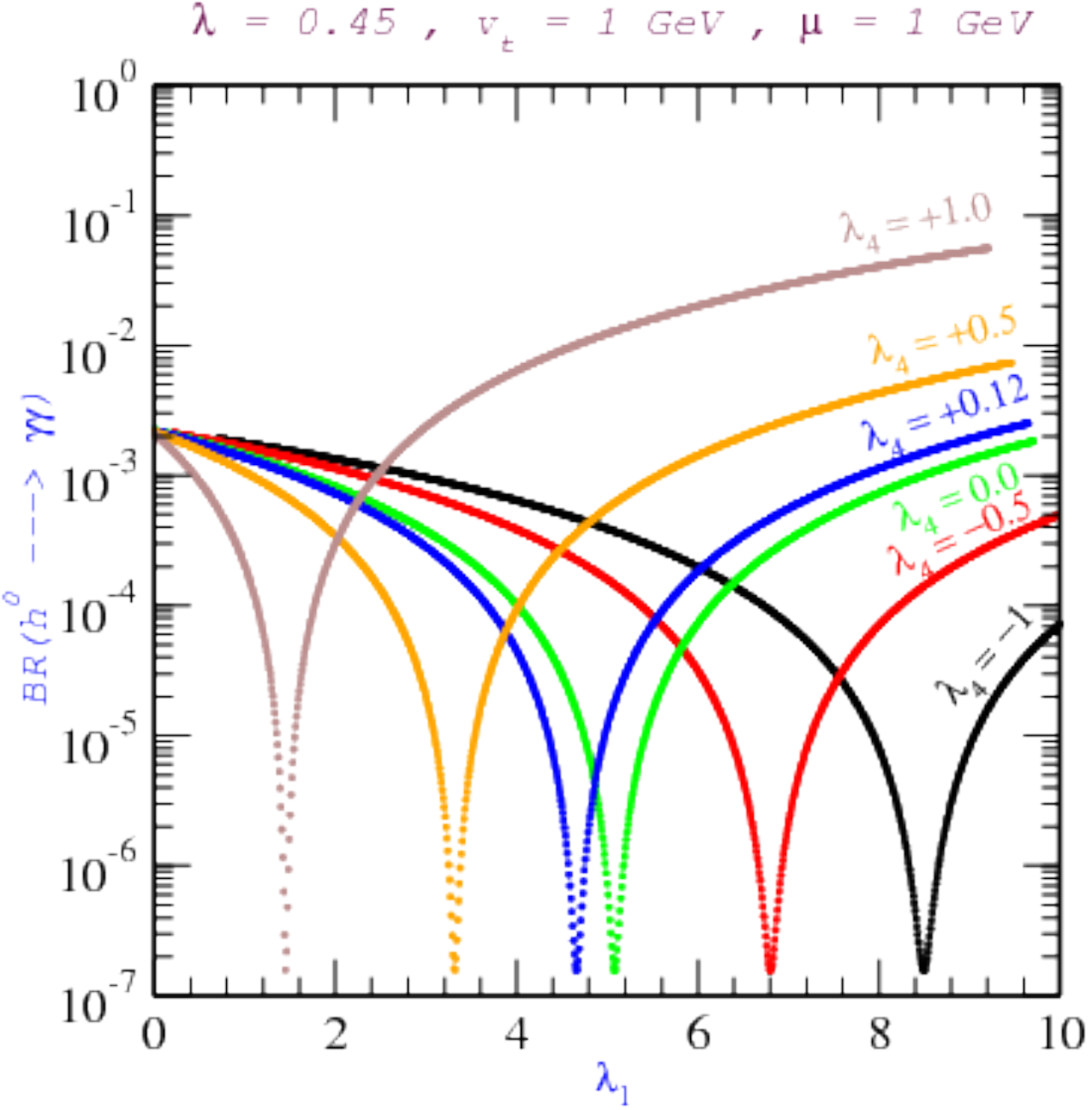}}
\vspace{0.4cm}
\centerline{\includegraphics[width=0.45\columnwidth]{sat33.eps}}
\caption{upper panel:
$Br(h\rightarrow \gamma\gamma)$ as
a function of $\lambda_1$ for various values of $\lambda_4$ and $\lambda=0.45$
corresponding to $m_{h}\approx 115$ GeV.
Lower panel: Scatter plot  in the $(\lambda_1,\lambda_4)$ plane showing
 $Br(h\rightarrow \gamma\gamma$.
In both panel: $\lambda=0.55$, $\lambda_3=2\lambda_2=0.2$ and $v_t=\mu=1$~GeV,
$h$ is SM-like and $m_{h}\approx 127$~GeV.
}\label{fig1}
\end{wrapfigure}

We show in Fig.~\ref{fig1} (upper panel) the $Br(h\to \gamma\gamma)$
as a function of $\lambda_1$, illustrated
for several values of $\lambda_4$ and $\lambda=0.45$, $v_t=1$ GeV.
In this plot, the lightest CP-even state $h$ carries $99$\% of
the SM-like Higgs component, with an essentially fixed mass
$m_{h}\approx 114$--$115$~GeV over the full range of values
considered for  $\lambda_1$ and $\lambda_4$.

%

As can be seen from this plot, ${\rm Br}(h\to \gamma\gamma)$ is very
close to the SM prediction [$\approx 2\times 10^{-3}$] for small
values of $\lambda_1$, irrespective of the values of $\lambda_4$.
Indeed in this region the di-photon decay is dominated by the SM
contributions, the $H^{\pm \pm}$ contribution being shutdown for
vanishing  $\lambda_1$, while
the sensitivity to $\lambda_4$ in the $H^\pm$ contribution,
is suppressed by a large $m_{H^\pm}$ mass, $m_{H^\pm}\approx 164$--$237$~GeV
for $-1 <  \lambda_4 <1$, cf. Eq.(\ref{eq:gcalHHpp}).
Increasing $\lambda_1$ (for fixed $\lambda_4$) enhances
the $g_{hH^\pm H^\mp}$ and $g_{hH^{\pm\pm} H^{\mp\mp}}$ couplings.
The destructive interference between the SM loop contributions and those of
$H^\pm$ and $H^{\pm \pm}$ becomes then more and more pronounced.
The leading DTHM effect is mainly due to the $H^{\pm \pm}$ contribution,
the latter being enhanced with respect to $H^\pm$ by a factor $4$
due to the doubled electric charge, but also  due to a smaller mass
 than the latter in some parts of the parameter space,
$m_{H^{\pm \pm}} \approx 110$--$266$~GeV.
It is obvious that the amplitude for $h\to \gamma\gamma$
is essentially linear in $\lambda_1$, since $m_{H^\pm}$ and $m_{H^{\pm\pm}}$,
Eqs.~(\ref{eq:mHpm}, \ref{eq:mHpmpm}), do not depend on
$\lambda_1$ while the dependence on this coupling through
$m_{h}$ is screened by the mild behavior of the scalar
functions  $A_{0, 1/2, 1}^{h}$. Furthermore,
the latter functions remain real-valued in the considered
domain of Higgs masses.
There exists thus necessarily values of $\lambda_1$ where the
effect of the destructive interference
is maximized leading to a tremendous reduction of
$\Gamma(h\to \gamma\gamma)$.
Since all the other decay channels remain SM-like, the same reduction occurs
for ${\rm Br}(h\to \gamma\gamma)$.
 The different dips seen in Fig.~\ref{fig1} are due to such a severe
cancelation between SM loops and $H^\pm$ and $H^{\pm\pm}$ loops,
and they occur for $\lambda_1$ values within the allowed
unitarity \& BFB regions. Increasing $\lambda_1$
beyond the dip values, the contributions of
$H^{\pm\pm}$ and $H^{\pm}$  become bigger than the SM contributions
 and eventually come to largely dominate for sufficiently
large $\lambda_1$. There is however another interesting effect
when $\lambda_4$ increases. Of course the locations of the dips depend also on
the values of $\lambda_4$, moving them to lower values of  $\lambda_1$
 for larger $\lambda_4$. Thus, for larger $\lambda_4$, there is a place,
within the considered range of $\lambda_1$, for a significant
increase of ${\rm Br}(h\to \gamma\gamma)$ by even
more than one order of magnitude with respect to the SM prediction.
This spectacular enhancement is due to the fact that larger $\lambda_4$
leads to smaller $H^{\pm\pm}$ and $H^\pm$ which can efficiently
boost the reduced couplings that scale like the inverse second
 power of these masses.
For instance varying $\lambda_4$ between $-1$ and $1$ in the upper panel case,
decreases $H^{\pm\pm}$ from $266$ to $110$ GeV, which modify
 the {\rm Br}($h\to \gamma\gamma$) by 2 orders of magnitude
with respect to the SM value.

In Fig.~\ref{fig1}, we show a scatter plot for
Br($h\to \gamma\gamma$) in the $(\lambda_1,\lambda_4)$ plane
illustrating more generally
the previously discussed behavior, for $m_{h} = 127$~GeV, imposing
unitarity and BFB constraints as well as the lower bounds
$m_{H^\pm} \gtrsim 80$~GeV and $m_{H^{\pm \pm}} \gtrsim 110$~GeV on the
(doubly-)charged Higgs masses\footnote{
Recently, CMS puts a lower limit of 313 GeV on $H^{\pm\pm}$ from $H^{\pm\pm}$
decaying leptonically. This limit can be reduced down to 100 GeV if one takes
into account the decay channels $H^{\pm\pm}\to W^{\pm} W^{\pm *}$ as well as
$H^{\pm\pm}\to H^{\pm} W^{\pm *}$ \cite{Melfo:2011nx,Perez:2008ha}.}.
One retrieves the gradual enhancement of Br($h\to \gamma\gamma$)
in the regions with large and positive
$\lambda_{1,4}$. The largest region (in yellow) corresponding to
${\rm Br}(h\to \gamma\gamma)\la 2\times 10^{-3}$
encompasses three cases: --the SM dominates
--complete cancelation between SM and $H^\pm$, $H^{\pm\pm}$ loops --$H^\pm$,
$H^{\pm\pm}$ loops dominate but still leading to a SM-like branching ratio.


Besides the branching ratio of $h\to \gamma\gamma$,
 and in order to compare our predictions with CMS and ATLAS data,
we will consider  the following observable relevant for LHC:
\begin{equation}
R_{\gamma\gamma}(h)=\frac{(\Gamma(h\rightarrow gg)
\times {\rm Br}(h\rightarrow \gamma\gamma))^{DTHM}}
{(\Gamma(h\rightarrow gg)\times
{\rm Br}(h \rightarrow \gamma\gamma))^{SM}}
\label{eq:Rgg}
\end{equation}
The above ratio has the advantage  that all
the leading QCD corrections as well as PDF uncertainties drop out.
Moreover, $R_{\gamma\gamma}$ can be viewed as an estimate of the ratio of
DTHM to SM of the  gluon fusion Higgs production cross section
with a Higgs decaying into a photon pair.
One should, however, keep in mind the involved approximations:
assuming only one intermediate (Higgs) state, one should take
the ratio of the parton-level cross-sections
$\sigma( g g \rightarrow \gamma \gamma)$ in both models, which are given
by $ {\rm Br}(h \rightarrow g g) \times
{\rm Br}(h \rightarrow \gamma\gamma)$. Using instead the
ratio $R_{\gamma\gamma}$ as defined in Eq.~(\ref{eq:Rgg})
relies on the fact that in the SM-like Higgs regime of DTHM,
the branching ratios of all Higgs decay channels are the {\sl same}
as in the SM, except for  $h\rightarrow \gamma\gamma$
 (and $h\rightarrow \gamma Z, gg$)
where they can significantly differ, but remain very small compared to the other
decay channels, so that $\Gamma(h \rightarrow {\rm all})^{DTHM}/
\Gamma(h \rightarrow {\rm all})^{SM} \approx 1$.

\begin{wrapfigure}{r}{0.5\columnwidth}
\centerline{\includegraphics[width=0.43\columnwidth]{FIG4-a.eps}}
\vspace{0.4cm}
\centerline{\includegraphics[width=0.43\columnwidth]{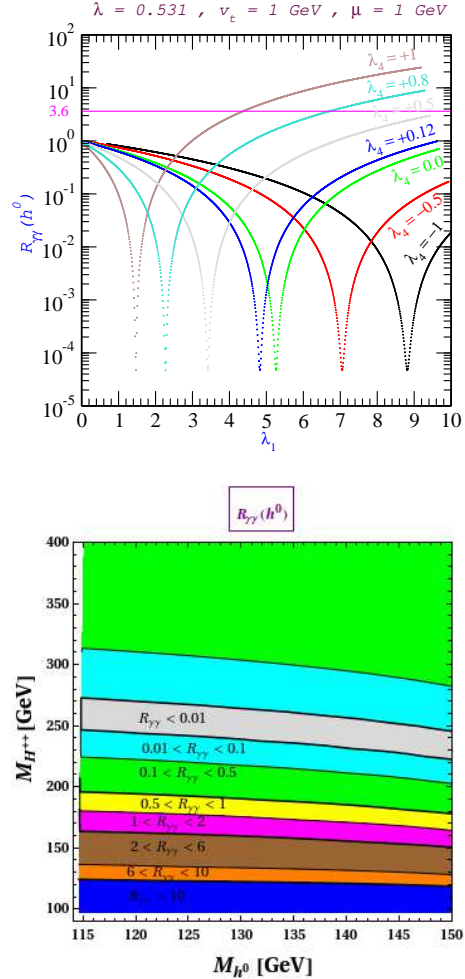}}
\caption{Upper panel: $R_{\gamma\gamma}$ as a function of
$\lambda_1$ for various values of $\lambda_4$,
with $\lambda=0.53$ ($h$ is SM-like and $m_{h}=124$--$125$~GeV),
$\lambda_3=2\lambda_2=0.2$ and $v_t=1$ GeV;
The horizontal lines indicate the ATLAS exclusion limits
\cite{ATLAS-CONF-2011-161}.
Lower panel: Scatter plots in the ($m_{h}, m_{H^{\pm \pm}}$) plane,
showing domains of  $R_{\gamma\gamma}$ values, we scan in the domain
 $.45 < \lambda < 1.2, -5 < \lambda_4 < 3$ with
$\lambda_1=8$, $\lambda_3=2\lambda_2=0.2$ and $v_t=\mu=1$ GeV.}
\label{fig2}
\end{wrapfigure}

In Figs.~\ref{fig2} we illustrate the effects directly in terms of the ratio
$R_{\gamma \gamma} \approx \sigma^{\gamma \gamma}/\sigma_{\rm SM}^{\gamma
\gamma}$ defined in Eq.(\ref{eq:Rgg}). We also show on the
upper plot Figs.~\ref{fig2} the present experimental exclusion
limits corresponding to these masses, taken
from \cite{ATLAS-CONF-2011-161}. As can be seen from Fig.\ref{fig2},
one can easily accommodate, for $m_{h} \approx 125 {\rm GeV}$,
a SM cross-section, $R_{\gamma \gamma}(m_{h}=125 {\rm GeV})=1$,
or a cross-section in excess of the SM, e.g.
$R_{\gamma \gamma}(m_{h}=125 {\rm GeV}) \sim 3$--$4$,
for values of $\lambda_1, \lambda_4$ within the theoretically
allowed region. The excess reported by ATLAS and CMS in
the diphoton channel can be readily
interpreted in this context. However, one should keep in
mind that all other channels remain
SM-like, so that the milder excess observed in
$WW^*$ and $ZZ^*$ should disappear with higher
statistics in this scenario. This holds independently
of which of the two states, $h$ or $h$, is playing the role of
the SM-like Higgs.
We comment now on another scenario, in case the reported excess around
$m_{h} \approx 125$~GeV would not stand the future accumulated statistics.
Fig.\ref{fig1}(upper panel) shows different dips in $Br(h\to \gamma \gamma)$
corresponding to the case of  $m_{h}\approx 115$~GeV. The ratio
$R_{\gamma\gamma}$ would have similar behavior as reported in
Fig.\ref{fig1}(upper panel) since $\Gamma(h\to gg)$
will be quite similar both in SM and DTHM.
Then, the large deficit for $R_{\gamma \gamma}$ in parts of
the $(\lambda_1, \lambda_4)$ parameter space opens up an unusual possibility:
the exclusion of a SM-like Higgs, such as the
one reported by ATLAS in the $114$--$115$~GeV range,
does not exclude the LEP events as being real SM-like Higgs
events in the same mass range!
This is a direct consequence of the fact that in the model we
consider, even a tremendous reduction in
$\sigma^{\gamma \gamma}=\sigma^{h} \times
{\rm Br}(h \to \gamma \gamma)$
leaves all other channels, and in particular the LEP relevant cross-section
$\sigma(e^+ e^- \to Z h)$ essentially identical to that of the SM.

Last but not least,  exclusion limits or a signal in the diphoton
channel can be translated into constraints on the masses of
$H^{\pm \pm}$ and $H^\pm$. We show in Figs.~\ref{fig2}(lower panel)
the correlation between $m_{h}$ and $m_{H^{\pm \pm}}$ for different ranges of
$R_{\gamma \gamma}$. Obviously, the main dependence on
$m_{h}$ drops out in the ratio $R_{\gamma \gamma}$
 hence the almost horizontal bands in the plots.
There remains however small correlations which are due to
the model-dependent  relations between the (doubly-)charged
and neutral Higgs masses that can even be magnified in the
regime of $h$ SM-like. In this plot we take large
value for $\lambda_1=8$ which give $R_{\gamma \gamma}>1$ for light
$m_{H^{\pm\pm}}$. For low values of $\lambda_1$, as we learn from previous
discussion, the ratio $R_{\gamma \gamma}$ remains below $1$ even for
increasing $H^{\pm \pm}$ and $H^\pm$ masses.
 The reason is that these masses become large when
$\lambda_4$ is large (and negative)  for which the loop
 contribution of $H^\pm$ does not vanish, as can be easily seen
 from Eqs.~(\ref{eq:mHpm}, \ref{eq:gcalHHp}).

\section{Discussion and Conclusions }
The lightest neutral CP-even Higgs boson $h$ of DTHM,
has the same tree-level couplings to
the SM weak bosons $Z$ and $W^\pm$ , leptons
and quarks, as the SM Higgs boson.
For these reasons, the main production mode for such a Higgs is the
usual gluon fusion, W fusion or Higgsstrahlung processes at
LHC or Higgsstrahlung and W fusion  processes at ILC.
As a matter of fact, the cross sections from all
production modes are expected to be of the same size as the SM Higgs.
Moreover, Higgs boson $h$ should have the same branching ratios to those SM
particles to which it couples at tree level as the SM Higgs except for a small
 suppression due to the mixing of the doublet with the triplet component.
The loop-induced decays, however, receive contributions from the charged states
of the model. This is indeed the case for the decays $h\to \gamma\gamma,
\gamma Z$  but not for $h\to gg$.
Therefore, it is expected that the loop mediated processes
$h\to \gamma\gamma, \gamma Z$ would have some large deviation
from its SM values but the total width of SM-like Higgs boson are nearly
identical to the SM Higgs one.
Any deviations from the SM case concerning the
SM-like Higgs boson production and decay at the LHC and/or ILC
are expected to come solely from the  branching
ratio into two photons and not from the production cross sections
or decays into the other conventional channels unless if the sensitivity 
of those measurements are of comparable size with radiative corrections. 

To conclude, we have discussed a possible enhancement/suppression
of the partial decay width $\Gamma(h\to \gamma\gamma)$ in the
DTHM for light charged states $H^\pm$ and $H^{\pm\pm}$.
The partial decay width $\Gamma(h\to \gamma\gamma)$ depends on
the potential via the couplings $hH^\pm H^\mp$, $hH^{\pm\pm} H^{\mp\mp}$.
This means that a possible enhancement depends on the
parameter space of the scalar potential.
Restricting the parameter space of the DTHM
with perturbative unitarity as well as vacuum stability constraints
 would restricts the possible enhancement of $\Gamma(h\to \gamma\gamma)$. 
In large area of parameter space of the DTHM,
the deviations in the decay width for $h \to \gamma\gamma$
from the corresponding SM values can be quite large,
therefore $h\to \gamma\gamma$ can be used to distinguish
between DTHM and SM through the precise measurement program planed at ILC and
its $\gamma\gamma$ option.

\section{Acknowledgements}
A. A is grateful to the organizers for the financial support. The work of R.B was supported by the Spanish Consejo Superior de
Investigaciones Cientificas (CSIC).


\begin{footnotesize}

\end{footnotesize}
\end{document}